\documentclass[]{emulateapj}

\usepackage{amsmath}
\usepackage{xspace}
\usepackage{graphicx}
\usepackage{txfonts}
\usepackage{natbib}
\def\xspec{{\sc xspec}}

\def\ktbb{kT_{\rm bb}}
\def\ktw{kT_{\rm w}}

{\def\kte{kT_{\rm e}}

\def\comptt{{\sc comptt}}
\def\bmc{{\sc bmc}}

\def\gaussian{{\sc gaussian}}
\def\chiq{$\chi^2$}
\def\cm2{cm$^{-2}$}
\def\s1{s$^{-1}$}
\def\mdot{$\dot{M}$}

\def\wabs{{\sc wabs}}
\def\sax{{\it BeppoSAX}}
\def\xte{{\it RXTE}}
\def\integral{\mbox{{\it INTEGRAL}}}

\shorttitle{Bulk motion Comptonization in LMXBs}
\shortauthors{Farinelli, Titarchuk and Frontera}

\begin{document}

\title{The hard X-ray tails in neutron star low mass X-ray binaries: \sax\  observations and possible theoretical explanation of the \mbox{GX 17+2} case}

\author{Ruben Farinelli\altaffilmark{1}, Lev Titarchuk\altaffilmark{1,2,3} \&  
Filippo Frontera\altaffilmark{1,4}}
\altaffiltext{1}{Dipartimento di Fisica, Universit\'a di Ferrara, via Saragat 1, I--44100, Ferrara, Italy}
\altaffiltext{2}{George Mason University/CEOSR/US Naval Research Laboratory, USA}
\altaffiltext{3}{NASA/GSFC, code 661,  Greenbelt, MD 20771, USA}
\altaffiltext{4}{INAF, Sezione di Bologna, via Gobetti 1, I--40129, Bologna, Italy}

\begin{abstract}
We report  results of a new spectral analysis of two \sax\ observations of the Z source {\mbox GX 17+2}. In one of the two observations the source exhibits a powerlaw-like hard ($> 30$ keV) \mbox{X-ray} tail which
was described in a previous work by a hybrid Comptonization model. Recent high-energy observations with  \integral\ of a sample of Low Mass X-Ray Binaries including both  Z and atoll classes have shown that  bulk (dynamical) Comptonization of soft  photons can be a possible alternative  mechanism for producing hard X-ray tails in such systems. We start  from  the  \integral\  results and  we exploit the broad-band capability of \sax\ to better investigate the physical processes at work. We use \mbox{GX 17+2} as a representative case.  Moreover, we suggest that weakening (or disappearance) of the hard X-ray tail can be explained by  increasing radiation pressure originated at the surface of the neutron star (NS).  As a result the high radiation pressure stops   the bulk inflow and consequently   this radiation feedback of the NS  surface  leads to quenching   the bulk  Comptonization.
\end{abstract}

\keywords{stars: individual: {\mbox GX 17+2} --- stars: neutron ---  X-rays: binaries --- accretion, accretion disks}

\maketitle

\section{Introduction}
\label{introduction}

One of the most interesting features of the high-energy \mbox{X-ray} Astronomy since the \sax\ and \xte\ era is the observation of a transient powerlaw (PL)-like emission above 30 keV in low mass \mbox{X-ray} binaries (LMXBs) hosting a weakly magnetized neutron star (NS). This component, which appears over the persistent continuum composed by
a thermal Comptonization (TC) plus a soft blackbody (BB)-like emission, up to now has been observed in the six
\mbox{X-ray} sources belonging to the so-called Z class ({\mbox GX 17+2}, Di Salvo et al. 2000, hereafter DS00, Farinelli et al. 2005, hereafter F05; {\mbox GX 349+2}, Di Salvo et al. 2001; \mbox{GX 340+0}, Lavagetto et al. 2004; \mbox{Sco X--1}, D'Amico et al. 2001; \mbox{GX 5--1}, Asai et al. 1994, Paizis et al. 2005; {\mbox Cyg X--2}, Frontera et al. 1998, Di Salvo et al. 2002), in the peculiar source {\mbox Cir X-1} (Iaria et al. 2001, 2002) and, very recently by \integral\  (Paizis et al. 2006, hereafter P06), in the source {\mbox GX 13+1}.  It is worth noting, although {\mbox GX 13+1} is  classified as atoll source   by Hasinger \& van der Klis (1989),  shows, in fact,   spectral properties similar to Z sources.

The hard \mbox{X-ray} emission seems to be correlated with lower \mdot\ and radio-loud states (see, e.g., Penninx et al. 1988 for \mbox{Sco X--1} and {\mbox GX 17+2}), but its physical origin is still debated. 

The thermal Comptonization of soft photons in very high-temperature plasma does not seem to apply because
it is very difficult to explain where and how  such a hot plasma could be present in a source where
a spectral signature of a low-temperature ($\sim$ 3-5 keV)  plasma is definitely observed (see, e.g., review by Di Salvo \& Stella 2002).

The first attempt to describe the hard \mbox{X-ray} tail  in terms of some physical self-consistent model was performed by F05 who fitted the \sax\ broad-band spectrum of {\mbox GX 17+2} using a hybrid Comptonization model (Coppi 1999) plus a soft $\sim$ 0.6 keV BB-component.
The presence or absence of the hard tail was   explained by decreasing the power
(supplied by some unspecified mechanism) injected to accelerate some fraction of electrons over their Maxwellian energy distribution.   

Titarchuk, Lapidus \& Muslimov (1998), hereafter TLM98,  showed that the Keplerian disk flow should adjust to the sub-Keplerian rotation of the central object, NS or Black Hole (BH). For a given viscosity in the disk 
the location of the adjustment radius $R_{adj}$ depends on the mass accretion rate only.
Thus in the transition layer (TL)  between $R_{NS}$ and  $R_{adj}$  the rotational velocity is sub-Keplerian. Moreover  Titarchuk, \& Farinelli (2007 in preparation), hereafter TF07, calculate the radial velocity $v_{R}$  in the TL and they find that $v_{R}$ is very close to the sound speed $v_{S}$ in the zone between $R_{adj}$ and some radius $R_{ff}$ below which the accretion onto NS occurs in almost  free-fall manner.

 Thus   TF07 unambiguously show the  bulk inflow must be considered a fundamental process in accretion into the compact objects.
Titarchuk, Mastichiadis \& Kylafis (1996), hereafter TMK96, studied  the effect of  the bulk motion Comptonization (BMC) of soft photons by energetic electrons in  a converging  flow into NS.  They found that the BMC emergent spectrum has a specific hard tail which  can be a possible explanation for  the hard \mbox{X-ray} emission in NS sources.

It is worthwhile to emphasize that this BMC  process was not  considered by F05  a possible mechanism to produce the observed hard tail of X-ray spectra from   NS sources.   

However  P06 analyzed  high-energy ($>20$ keV) observations of NS systems performed by \integral\  and they confirmed that BMC indeed can explain this  hard \mbox{X-ray} emission.
 
P06 supported this spectral evolution picture of NS sources using a sample  of twelve bright low-mass \mbox{X-ray} binaries hosting a NS. Their  sample comprises the six  Galactic Z sources and six atoll sources, four of which are bright GX bulge sources while two ({\mbox H~1705--440} and {\mbox H~1608--552}) are weaker  in the 2$-$10 keV range. 

Comparing their results with those obtained by Falanga et al. (2006) from \mbox{GX 354--0} (4U~1728--34), P06 identified four main spectral states for NS LMXBs (see Fig. 4 in P06): low/hard state (\mbox{GX 354--0}), hard/PL state  (\mbox{H~1705--440} and \mbox{H~1608--552}), intermediate state (e.g., \mbox{GX 5--1}) and very soft state (e.g., GX 3+1).
Above 20 keV, the low/hard spectra are described by TC with  plasma temperature $\kte \sim$ 30 keV and optical depth  $\tau \sim$ 1;
 the hard/PL state spectra, as suggested by the name, show simple PL-like emission; the  intermediate state spectra are characterized by the presence of a TC-component with spectral parameters similar to those of the soft state plus additional PL-like emission (above $\sim$ 30 keV); and the soft state spectra are described by a single TC-component  with low-$\kte$  and high-$\tau$.

We remind the reader that P06 used the high-energy ($>20$ keV) \integral\ observations of these NS sources, so they could not reveal the soft BB-component observed in NS LMXBs (see review by Barret 2001).
Using broad-band data, the intermediate states of NS (see, e.g., {\mbox GX 17+2} and {\mbox Cyg X-2}), being similar to high-soft states of BHCs, clearly show the presence of a soft BB-like component plus a PL emission at high energies, but with an additional TC feature that is absent in BHC spectra. This is because the  presence of a solid surface leads to a further channel of emission. Thus, this similarity of spectral appearances in NSs and BHCs suggests a similarity  in the physical processes occurring in these two classes of systems.

In this work, using the excellent broad-band observing capability of {\it BeppoSAX} (Boella et al. 1997a),
we present  the 0.4-120 keV energy spectrum of the Z source {\mbox GX 17+2}, in order to shed light on the source geometry.
We find that the BMC model can really explain the origin of the high-energy emission in NS LMXBs.

\section{The Bulk Comptonization Model}

The TC  theory  along with the dynamical (bulk motion) Comptonization  has been first  suggested
by Blandford \& Payne (1981, hereafter BP81), and further developed in detail by Titarchuk, Mastichiadis \& Kylafis (1997, hereafter TMK97).
BP81 derived the diffusion Fokker-Planck equation using the sub-relativistic kinetic equation 
[see Eq. [15] in BP81 and Eq. [14] in TMK97].
The relative efficiency of TC effect with respect to BMC is  
determined as (see TMK97) 
\begin{equation}
\frac{<\Delta E_{th}>}{<\Delta E_{b}>}
 < \frac{1}{\delta},
\label{ratio}
\end{equation}
where $<\Delta E_{th}>$ and $<\Delta E_{b}>$ represent the average energy photon exchange due to TC and BMC respectively. The $\delta$ parameter is defined as (see TMK96 and TMK97)
\begin{eqnarray}
\delta \equiv {(1-\ell)}^{0.5}/(\dot m\Theta) 
=51.1\times {(1-\ell)}^{0.5} T_{10}^{-1}\dot m^{-1},
\label{delta}
\end{eqnarray}
where $\Theta \equiv \kte/m_ec^2$ is a dimensionless plasma temperature and $T_{10}\equiv \kte/(10 $ keV), while $\ell \equiv L/L_{\rm Edd}$ and $\dot{m} \equiv \dot{M}/\dot{M}_{\rm Edd}$ are a dimensionless luminosity emitted by the NS and  mass accretion rate calculated in Eddington units
($\dot{M}_{\rm Edd}=L_{\rm Edd}/c^2$)  respectively.
When $\delta=0$, diffusion equation (14) of TMK97 is reduced to the TC diffusion equation (see, e.g., Titarchuk 1994). For our purposes it is just sufficient to point out that the photon occupation number of the emergent spectrum can be expressed as
\begin{equation}
F(x)=\sum_{k=1}^{\infty} \int_0^{\infty}I_k(x,x_0)s(x_0)dx_0=\sum_{k=1}^{\infty} I_k(x,x_0)\ast s(x_0),
\label{convolution}
\end{equation}
where $x \equiv E/\kte$, $x_0 \equiv E/\ktbb$,  $\kte$,  $\ktbb$ are  temperatures of  plasma and of seed photons, respectively and $\ast$ is the convolution operator.

 TMK97  showed that $I_1(x,x_0) \propto x^{-\alpha}$   for $k=1$ and $x\geq x_0$ up to a rollover energy of order of $E_c \approx {m_ec^2 / \dot{m}}$, while $I_k(x,x_0) \propto \delta(x-x_0)$ 
  for $k\geq 2$, so that 
$I_k\ast s\propto\int_0^{\infty}\delta(x-x_0)s(x_0)dx_0 \approx s(x)$.
In other words, only the first  term ($k=1$) in the series of Eq. [\ref{convolution}] significantly contributes to the upscattering (Comptonization) part of the emergent  spectrum. The terms  related to  $k\geq2$ represent the photons
which escape the plasma cloud with insignificant energy-change (see also Fig. 3 in TMK97) and thus they retain the shape the original seed photon spectrum.

The relative importance of the Comptonization term $I_1\ast s$ with respect to that of the
seed photons depends on their spatial distribution  within the cloud
(see Fig. 4 in TMK97).
In the BMC model for \xspec\ developed by TMK97, the  seed photon spectrum $s(x_0)$ of Eq. [\ref{convolution}] is assumed to be a BB-like spectrum, while the relative importance of the efficiently  up-scattered photons, the term $I_1\ast BB$ of series (\ref{convolution})  with respect to the direct BB component  is expressed using a weighting factor $A/(A+1)$. In fact, the \xspec BMC model reports $\log (A)$. Thus the analytical formula (back-reverting from $x$ to $E$) for the total spectrum is
\begin{equation}
F(E)= \frac{ C_N}{A+1} (BB + A \cdot I_1\ast BB).
\label{BMCmodel}
\end{equation}
The free parameters for \xspec\ are  temperature of the seed photons $\ktbb$,  spectral index $\alpha$ of the term $I_1$, a logarithm of the weighting factor $\log(A)$ and  the normalization $C_N$.
The spectral index $\alpha$  is related to the efficiency of the Comptonization processed (both thermal and bulk). As higher  $\alpha-$value, the Comptonization efficiency is less and thus  the system is  closer  to thermal equilibrium, so that the emergent spectrum has  a BB-like shape.
The  weighting factor $A$ is chosen in such a way  that for $A \rightarrow 0$ the model is  reduced to the standard BB model of \xspec.

\section{Results}
\label{results}

Using \xspec\ v11.3  we made a spectral analysis of a set of two \sax\  observations of the Z source {\mbox GX 17+2}  performed in 1997  (on April 3 and 21, respectively).  A detailed description of the data processing and analysis can be found in F05. We just point out that in both cases the source was in the horizontal branch (HB) of its colour-intensity diagram (HID).  As during the first observation the source  stayed in a clustered region of  the HB, whereas in the second observations the source evolved along the branch, from right to left (see Fig. 2 in F05). The HB region of the second observation was divided into three parts, for which separated spectral analysis was carried out. Given that there was only a very slight variation of the spectral parameters in these regions, F05 reported their results in terms of the time-averaged spectrum.

In the {\it first} observation  F05 found a systematic excess in the residuals above 30 keV  by  fitting the persistent continuum  with  a photoelectrically-absorbed (\wabs\ in \xspec) BB plus the TC model of Titarchuk (1994, \comptt\ in \xspec). This excess was {\it phenomenologically} described by a PL with photon index $\Gamma \sim 2.8$. DS00  obtained  the same $\Gamma$ value for a  different set of \sax\ observations, when the source was in the left part of the HB. 

 Given that the output shape of  the {\it physical} BMC  spectral model consists of  a direct (non up-scattered) BB-like component  along  with a PL  shape at high energies,  our new fitting model was \comptt\ plus \bmc\ plus a \gaussian\ emission line around 6.7 keV, which was previously observed both by DS00  and F05.

To vary all of the model parameters, we find that the data can be  presented  by two solutions in the framework of the {\it same} model.  One solution  is 
with $ \ktbb < \ktw $ (Case A)  and another one is  with $\ktbb > \ktw$ (Case B).  We remind a reader that 
$\ktbb$ and  $\ktw$  are seed photon temperatures for \bmc\ and \comptt\ models respectively.

We note that this dichotomy was already reported by F05 for the BB plus \comptt\ model. In both cases we obtain \chiq/dof$=151/144$. It implies that  we cannot statistically distinguish between these two solutions. 
In Case A we find  $\ktbb=0.62\pm 0.02$ keV, $\ktw=1.14\pm 0. 05$ keV,  $\log(A)=0.05~(> -0.66)$,  and 
$\alpha=2.1^{+   0.3}_{-   0.5}$, while for Case B the best-fit parameter values are  reported in Table \ref{fit} (see the first column).
It is worth noting that both the \comptt\ and \gaussian\ best-fit parameters are almost the same within error bars for  these two solutions.

In the {\it second} observation, the source was detected ($S/N \geq 3$) up to 50 keV.
A simple BB plus \comptt\ plus \gaussian\ model provides \chiq/dof=192/142 and \chiq/dof=197/142 for Case A ($\ktbb < \ktw$) and B ($\ktbb > \ktw$) respectively. As already discussed in F05, the reason for these non excellent \chiq\ values can be attributed to averaging  different (even though very close each other) spectral states. Indeed, no systematic deviations are observed in the residuals, and the last PDS point (40$\pm$ 10 keV) is at $\sim$2.5 $\sigma$ level above the model.
As a consequence of this, the \bmc\ parameters $\alpha$ and $\log(A)$ remains largely undetermined if they are left free in the fit.  We thus fixed $\alpha$ to 2.1 and 2.6 for Cases A and B respectively. In both Cases the fit quality only slightly improves,  the corresponding \chiq/dof decreases to value of 187/140.

A comparison  between the two observations, shows  that the major change is related to $\log(A)$, whose value decreases from  $\sim 0.73$ in the first observation to $\sim -0.33$ in the second  observation. This is consistent with the fact that the PL-like emission is significantly detected only in the first observation (see Table 1).

\begin{figure}[!th]
\centering
\includegraphics[width=7cm, height=9cm, angle=-90]{f1a.eps}
\includegraphics[width=7cm, height=9cm, angle=-90]{f1b.eps}
\includegraphics[width=16cm, height=8cm]{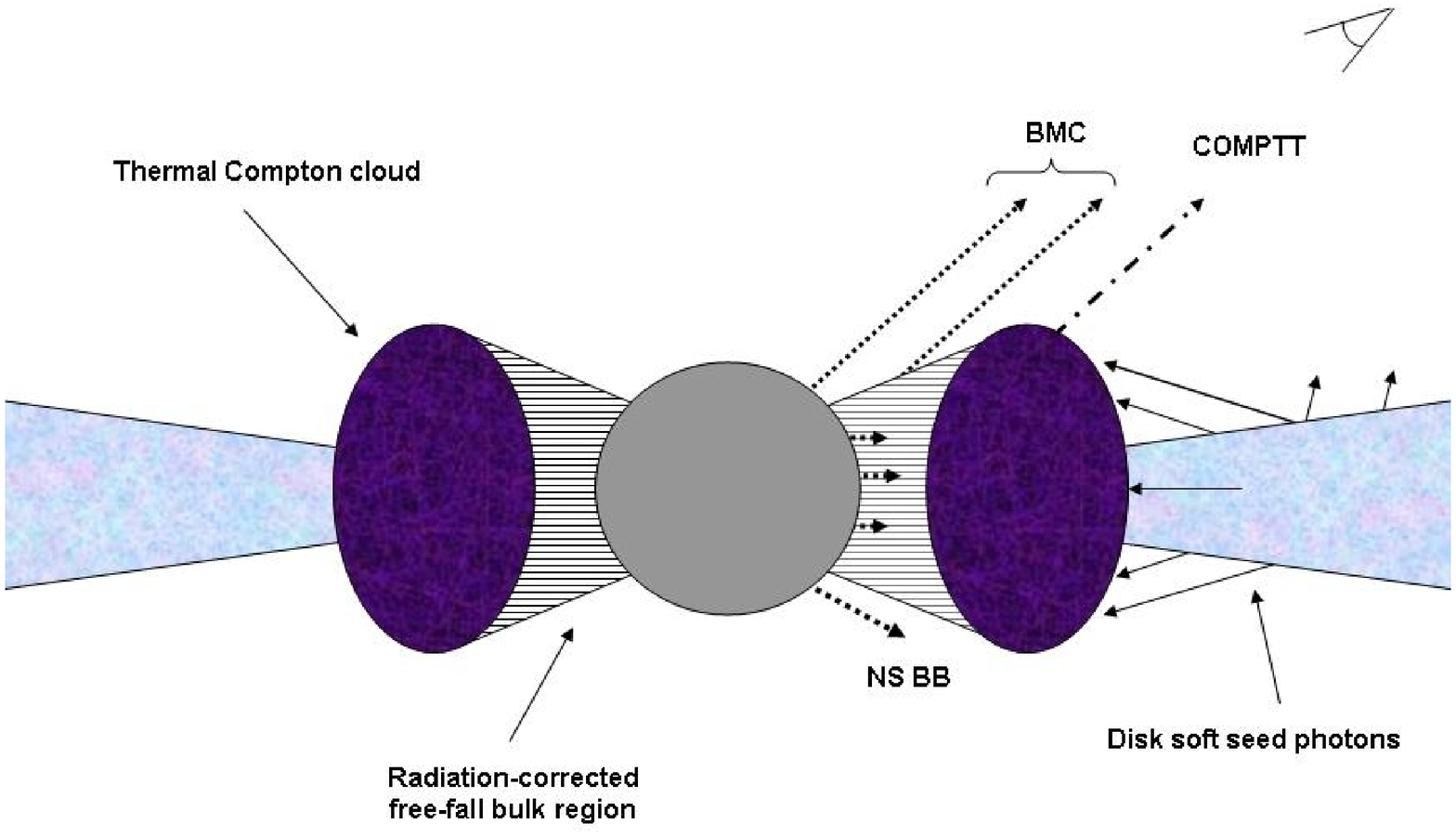}
\caption{{\it Top}: Unabsorbed EF(E) spectra and best-fit model \wabs(\bmc\ + \comptt\ + \gaussian) for the first ({\it upper panel}) and the second  ({\it lower panel}) observation of {\mbox GX 17+2} performed by \sax.    The residuals (in units of 1$\sigma$) between the data and the model
are also shown below each spectrum. Different line styles represent the single components. {\it Dotted-dashed}: \comptt. {\it Dotted}: \bmc.  {\it Long-dashed}: \gaussian. 
{\it Bottom}: Schematic view of the accretion geometry of GX 17+2 as inferred from the single spectral components observed in its X-ray spectrum (see {\it Top}). The decreasing (or disappearing) of the PL-like emission at high energies may be due to a decrease of  the substended solid angle of the radiation-corrected free-fall bulk region as  seen from the NS, or to higher  radiation pressure.  
It is also possible that both effects lead to the hard tail quenching.}
\label{efe}
\end{figure}

Below  we present  a physical reasoning  why  the best-fit parameters of  Case B are  more preferable  than that for Case A.  
In  Fig. \ref{efe} (top panel) we present  the deconvolved EF(E) spectra of the two observations for Case B to  illustrate this effect of the transient hard \mbox{X-ray} emission. 

\begin{table}
  \caption[]{Best-fit parameters of the multi-component model \wabs(\bmc\ + \comptt\ + \gaussian). Errors are computed at 90\% confidence level for a single parameter.}

\begin{tabular}{ccc}
\hline
\hline
\noalign{\smallskip}
Parameter          & Obs. 1		   &	Obs. 2\\

\noalign{\smallskip}
\hline

\noalign{\smallskip}

$N_{\rm H}^{\rm a}$ &  1.93$^{+   0.07}_{-   0.07}$  & 1.89$^{+   0.09}_{-   0.08}$ \\
\noalign{\smallskip}
\hline
\noalign{\smallskip}

\bmc\ &    & \\
\noalign{\smallskip}

$kT_{bb}$ (keV) & 1.33$^{+   0.14}_{-   0.10}$  & 1.31$^{+   0.07}_{-   0.08}$  \\

\noalign{\smallskip}
log(A)  & 0.73 $(> -0.24)$   & -0.33$^{+   0.29}_{-   0.39}$  \\
\noalign{\smallskip}
$\alpha$  &  2.64$^{+   0.19}_{-   0.56}$  &   [2.64]  \\

\noalign{\smallskip}
\hline
\noalign{\smallskip}
\comptt\ &    & \\
\noalign{\smallskip}

$kT_{\rm w}$ (keV) &  0.54$^{+   0.02}_{-   0.03}$ &   0.55$^{+   0.03}_{-   0.03}$ \\
\noalign{\smallskip}

$kT_{\rm e}$ (keV) & 3.32$^{+   0.07}_{-   0.06}$  &  3.11$^{+   0.09}_{-   0.08}$ \\
\noalign{\smallskip}
$\tau$ & 12.5$^{+   0.4}_{-   0.4}$  &  13.1$^{+   0.6}_{-   0.5}$ \\
\noalign{\smallskip}
\hline
\noalign{\smallskip}

\gaussian\ &    & \\
\noalign{\smallskip}

$E_{\rm l}$ (keV) & 6.69$^{+0.05}_{-0.05}$ &  6.72$^{+   0.08}_{-   0.09}$ \\
\noalign{\smallskip}
$\sigma_{\rm l}$ (keV) &   $0.21^{+0.09}_{-0.09}$    & 0.24$^{+0.13}_{-0.12}$   \\
\noalign{\smallskip}
$I_{\rm l}^{b}$ &  $5.2^{+1.1}_{-1.0}$  & $ 5.1^{+1.6}_{-1.3}$    \\
\noalign{\smallskip}
$EW_{\rm l}$ (eV)  &   38$^{+6}_{-9} $ & 35$^{+13}_{-8} $     \\

\noalign{\smallskip}
\noalign{\smallskip}
\hline
\noalign{\smallskip}
$F_{\rm bb}/F_{\rm tot}^{\rm c}$ & 0.16 & 0.21\\
\noalign{\smallskip}
$L_x^{\rm c,d}$ & 1.46 & 1.47 \\
\noalign{\smallskip}
${\chi^{2}}$/dof & 154/145 &  185/141   \\

\noalign{\smallskip}
\hline
\noalign{\smallskip}
\multicolumn{3}{l}{$^{\rm a}$ In units of 10$^{22}$ cm$^{-2}$.} \\
\multicolumn{3}{l}{$^{\rm b}$ Total photons in the line in units of 10$^{-3}$ \cm2 \s1.} \\
\multicolumn{3}{l}{$^{\rm c}$ Extrapolated in the energy range 0.1-200 keV.}\\
\multicolumn{3}{l}{$^{\rm d}$ In units of 10$^{38}$ erg s$^{-1}$ assuming a distance of 7.5 kpc}\\
\multicolumn{3}{l}{$$ (Penninx et al. 1988).}\\

\noalign{\vskip -0.cm}
\label{fit}

\end{tabular}
\label{fits}
\end{table}

\newpage

\section{Discussion and Conclusions}
\label{discussion}
In this Paper   we show the results of  a new \mbox{X-ray} spectral analysis, using \sax\ broad-band data,
of the Z source {\mbox GX 17+2} during two states in which a PL-like hard \mbox{X-ray} tail was present and absent, respectively. We describe this hard tail in the framework of Comptonization of soft BB-like seed
photons in the converging flow into NS.

Comparison of our best-fit parameters with those reported in P06 reveals  some significant
differences: Restricting energy-band to above 20 keV allows one to see only the thermal 
(BB-like or Wien-like) bump plus the additional PL-like \mbox{X-ray} tail in the spectrum (see, e.g., Fig. 5 in P06).
 This thermal  bump, lacking low-energy information, is interpreted by \bmc\  as the direct seed photon  component, namely the component which consists of the photons subjected   insignificant energy change when  they scatter in the ambient medium.
 This is confirmed by the fact that the color temperature $\ktbb$ of this component about $\sim$ 2.7 keV 
 (see Table 2 of P06), is not far away from  the electron temperature of \comptt\ found by us ($\sim 3.3$ keV, see Table \ref{fit}). 
 
 However using broad-band data it is possible to see  an additional  direct seed thermal emission
 located at lower energies whose  thermal Comptonized part dominates in the 20-40 keV region
 (see Fig. \ref{efe}, top panel).
One unavoidable consequence of these considerations is also the different values of the illuminating factor  A [or $\log(A)$].

The broad-band analysis allows us  to elaborate a more detailed picture of the source properties. As shown in Sect. \ref{results}, we can identify two regions of seed photon temperatures, for which the temperature values  differ  from each other by a factor about two. In  Case A the hotter ($\sim$ 1.2 keV) photons are  a  source for  thermal Comptonization, while the cooler ones ($\sim$ 0.6 keV) are up-scattered to form the PL emission at high energies. In  Case B the opposite situation occurs. Given that we cannot statistically distinguish one model from   another, the models give the same \chiq\ value,  only considerations based on the physical characteristics of the emitting regions can help us to discriminate among them.

Several theoretical studies (e.g., TLM98; Popham \& Sunyaev 2001, hereafter  PS01) have shown that  when the geometrically thin accretion disk approaches the NS, the difference in the angular velocity of the disk  and the more slowly spinning accreting star gives  rise to a transition region, usually called the boundary layer (BL)  in PS01 or transition layer (TL) in TLM98.
PS01 argue that when   \mdot\ increases, the BL extends radially (reaching 2-3 $R_{NS}$), significantly increasing its contribution to the total emitted luminosity (even to 70\%). Moreover PS01 show that inside the BL, the plasma properties (e.g., temperature, density, pressure, radial velocity) are not constant but vary as a function of the distance from the NS surface. 

TF07 demonstrate that in the TL, the radial  velocity profile as a function of radius is almost constant from the adjustment  radius  $R_{adj}$ (defined as the outer TL radius)  to some lower radius  $R_{ff}$ where the constant velocity flow is followed by the free-fall  accretion flow,  affected by radiation pressure (see also Introduction section).  TF07 obtain this radial velocity profile as a solution of the radial momentum equation [ see Eq. [5] in  PS01] for which they use  a  solution of the angular momentum equation found by TLM98 (see also Titarchuk \& Osherovich 1999).  
The free-fall region of the TL  is likely  a place  where the bulk motion Comptonizarion (BMC)  spectrum is formed. 

The free-fall region is  the innermost part of the TL,  and consequently the seed photons subjected to BMC  mainly come from the NS surface.
The temperature of the NS photons $\ktbb $ is higher than that of the disk soft photons $\ktw$ because the effective area of NS surface $\sim 3R_{\rm S}$ ($R_{\rm S}$ is the Schwarzchild radius)  is more compact  than that of the Keplerian disk $\sim (10-15)R_{\rm S}$ (see Shrader \& Titarchuk 1999).
This is a physical (not statistical)  reason why  we choose, among the two   solutions,  the one for  which $\ktbb > \ktw$. In Fig. \ref{efe} (bottom panel)  we provide a schematic view of the possible source accretion geometry.

The conservation of energy provides a free-fall radial velocity given by 
\begin{equation}
v_{\rm ff}=c [(1-F/F_{\rm Edd}) (R_{\rm S}/R)]^{1/2},
\label{free_fall}
\end{equation}
 where $F$ is the {\it local} flux and 

\begin{equation}
F_{\rm Edd}= \displaystyle{{G M c m_H n_i \mu_i} \over {R^2\sigma_{\rm T} n_e}},
\label{flux_edd}
\end{equation}
is the {\it local Eddington flux}. In Eq. [\ref{flux_edd}], $\sigma_{\rm T}$ is the Thomson cross-section,  $M$ is the NS mass,  $c$ is the speed of light, $n_i$ and $u_i$ are the ion density number and molecular weight, respectively, while $n_e$ is the electron density.  
 As it follows  from  Eqs. [\ref{delta}] and  [\ref{free_fall}] $v_{\rm ff} \propto \delta$ .
When  the {\it local} flux approaches  the {\it local Eddington flux}, the factor $(1-F/F_{\rm Edd})^{1/2}$ goes to zero and consequently  $v_{\rm ff}$ goes to zero too. 
Thus the effect of the disappearance of the BMC  hard tail ($\delta \rightarrow 0$) can be naturally explained as a result of the feedback  from the NS surface when the NS emergent (local) radiation flux is very close to Eddington.

In Table \ref{fit} one can see that, while the total flux remains almost the same,
the contribution of the seed BB-component (suggested to come from the NS) increases from the first observation ($F_{\rm bb}/F_{\rm tot}=0.16$) to the second one ($F_{\rm bb}/F_{\rm tot}=0.21$).  Given that the BB-flux  flux  is proportional  to the total  \mdot, we can conclude that  \mdot,  is a  quantity which regulates   appearance and disappearance of the BMC hard tail in X-ray spectra from NS LMXBs. 

If the firm surface is absent, like in BHs, then for a given electron temperature of the converging bulk flow the index $\Gamma$ decreases  with \mdot\ to some value and then it  saturates because of the photon trapping effect in the converging flow (see  more details in Titarchuk \& Zannias 1998). But in the NS case, as we explain above, there is a feedback from the NS surface when \mdot\ is very high. The high radiation flux emerged from the NS surface quenches the bulk free-fall flow.    
However, we may not  exclude that the hard tail, instead of disappearing, just decreases its  intensity, going below the instrument detection threshold, because of a lower fraction of seed (NS) photons intercepted  by the converging flow ({\it geometrical effect}).
The amount of weakly  upscattered seed photons leaving the flow depends on  their spatial distribution within the flow itself (TMK97). The BB-like direct component is stronger for more uniform spatial distribution  of  seed photons than that  mainly concentrated at the bottom. 

The changes of the observed contributions coming from  the BB-like and the PL component may be equally provided either by a change of  the seed photon spatial distribution (a more uniform distribution would decrease the amount of efficiently up-scattered photons, {\it spatial effect}) or by a lower substended angle of the flow as seen from the NS ({\it geometrical effect}). 
Another important point of our discussion concerns the apparent mass accretion rate \mdot. Very often, changes in \mdot\ in a given sources are inferred by measuring changing on its {\it apparent} luminosity, using the relation \mdot$= F_x 4 \pi D^2/(\eta c^2)$. However, this approach may lead to controversy. 
In fact, Di Salvo et al. (2002) and DS00 observed the hard X-ray tail in {\mbox Cyg X--2} and \mbox{GX 17+2} when the sources were in the HB of their HIDs.  The apparent 0.1--100 keV luminosity $L_{app}$ evolves  during the source  motion along Z-track from the  HB to the normal branch (NB).  $L_{app}$ increases in  {\mbox Cyg X--2}  and it decreases  in  \mbox{GX 17+2} with $S_z$ (where the length parameter  $S_z$ along the Z-track  increases from HB to NB).

Moreover the NS sources belonging to the high-luminosity atoll class GX 3+1, {\mbox GX 9+1} and \mbox{GX 9+9} have stable X-ray spectra very close to Z sources but, up to now, no evidence of PL-like hard X-ray emission has been provided for them. This is in constrast with what expected, given that, despite their spectral similarity, these sources are less luminous than Z sources.
Additionally, our data analysis shows that the hard X-ray behaviour changes at the same {\it apparent} luminosity level (see Table \ref{fit}).
All these observable quantities, show that mapping of \mdot\ changes either within the same source, or among different sources must be treated very carefully. 

Thus we can suggest that the NS surface feedback on the BMC tail is not sensitive to  the {\it absolute} (disk plus TL) \mdot\  value, but it mostly depends  on the {\it local} radiation flux related to the $\dot{M}_{\rm TL}$   impinging  on the converging flow.
This value may significantly differ from the simple $L/(4\pi R^2)$ estimate (see also PS01).

From the results of the multi-component fit, we are also able to map the energy budget of
the system. We reveal that  a fraction of the low-temperature (0.55 keV) BB radiation in the total flux, which is presumably the disk contribution to the total flux $F_{disk}/F_{tot}$,  is only  $\sim 34$\%.  The larger fraction of the gravitational energy of the accreting material $\sim 66$\%  is released in the transition layer and at the NS surface.
We infer $F_{disk}/F_{tot}$ using  a fraction of the TC-component (\comptt) in the total flux  
$F_{\rm comp}/F_{tot}\sim78$\% and a value of the  Compton enhancement factor $E_{\rm comp}\sim 2.3$.  Note the values of  $F_{\rm comp}/F_{tot}$ and 
$E_{\rm comp}$ are almost the same for the two observations.  We remind a reader that 
$E_{\rm comp}$ is defined as  a ratio of the  flux of the Comptonized emerging spectrum $F_{\rm comp}$ with respect to the soft photon injected flux $F_{soft}$ (see Sunyaev \& Titarchuk 1980).  
We calculate $E_{\rm comp}$  using \comptt-model. Namely, the model flux with the best-fit \comptt-parameters gives us $F_{\rm comp}$ whereas the model flux with the best-fit $\ktw$, $\kte$, normalization  and  very small optical depth $\tau$ provides us $F_{soft}$\footnote[1]{Given that the average number of scatterings  for photons is $\propto \tau^2$, it is just sufficient to set $\tau\ll1$ in \comptt\  and obtain $F_{soft}$. }.

We want also to emphasize that our preference to explain  the transient hard X-ray tails in NS LMXBs in the framework of the BMC phenomenon is not related to the fact that this model better describes the data statistically. The hybrid Comptonization model by Coppi (1999) first
used  by F05 to explain the \mbox{GX 17+2} \sax\ broad-band spectrum  was  statistically acceptable too.

The main reason for our choice is that  the BMC process is just related to the dynamics of the accretion flow, based on the exact solution of dynamical equations,  and {\it does not require any particular tuning and any particular energetic configuration of the Comptonizing plasma (e.g. thermal plus non-thermal distribution)}. 
We thus cannot a-priori exclude that non-thermal electron distributions are present in such a systems 
(for instance in possible emitting jets), but the key point is that {\it they are not required to explain our data}.

In  BH sources, the BMC  phenomenon are likely  even more prominent than  in NSs (because of the absence  of a solid surface and thus the absence of  a strong radiation feedback effect on the accreting flow) and their effect can well   explain the transition from hard/low   to  high/soft states in  these systems.

We acknowledge the referee for his/her important questions and for the constructive suggestions which improved  the paper quality.

\clearpage

\end{document}